\documentclass[pra,superscriptaddress,twocolumn]{revtex4-1}
\usepackage[T1]{fontenc}
\usepackage{amsmath}
\usepackage{amssymb}
\usepackage{mathtools}
\usepackage[normalem]{ulem}
\usepackage{graphicx}
\usepackage{datetime}
\usepackage{color}
\usepackage{newtxtext,newtxmath}
\usepackage{microtype}
\usepackage{braket}
\usepackage{xr}
\usepackage{natbib}
\usepackage{hyperref}
\hypersetup{%
    bookmarksopen=false,
    bookmarksnumbered=true,
    pdfnewwindow=true,
    unicode=false,
    colorlinks=true,%
    citecolor=blue,
    linkcolor=black,
    urlcolor=blue,
    filecolor=blue
    }  
\newcommand{\NN}{\mathcal{N}}

\renewcommand{\P}{\mathcal{P}}
\newcommand{\R}{\mathbb{R}}
\newcommand{\mrm}[1]{\mathrm{#1}}
\newcommand{\eps}{\varepsilon}
\renewcommand{\i}{\mathrm{i}}
\DeclareMathOperator{\var}{var}
\DeclareMathOperator{\Skew}{skew}
\renewcommand{\vec}[1]{\boldsymbol{#1}}
\newcommand{\Dq}[1]{\tilde{D}_{#1}}
\newcommand{\aDq}[1]{\big\langle\tilde{D}_{#1}\big\rangle}
\newcommand{\vDq}[1]{\var\big(\tilde{D}_{#1}\big)}
\newcommand{\sDq}[1]{\Skew\big(\tilde{D}_{#1}\big)}
\begin{document}
	\newcommand{\figdir}{.}
	\newcommand{\freiburg}{Physikalisches Institut, Albert-Ludwigs-Universit\"{a}t-Freiburg, Hermann-Herder-Stra{\ss}e 3, D-79104, Freiburg, Germany}
	\newcommand{\usal}{Departamento de F\'isica Fundamental, Universidad de Salamanca, E-37008 Salamanca, Spain}
	\newcommand{\eucor}{EUCOR Centre for Quantum Science and Quantum Computing, Albert-Ludwigs-Universit\"{a}t Freiburg, Hermann-Herder-Stra{\ss}e 3, D-79104, Freiburg, Germany}
	
	\title{Chaos in the Bose-Hubbard model and random two-body Hamiltonians}
	\author{Lukas Pausch}
	\affiliation{\freiburg}
	\author{Edoardo G.\ Carnio}
	\affiliation{\freiburg}
	\affiliation{\eucor}
	\author{Andreas Buchleitner}
	\email[]{a.buchleitner@physik.uni-freiburg.de}
	\affiliation{\freiburg}
	\affiliation{\eucor}
	\author{Alberto Rodr\'iguez}
	\email[]{argon@usal.es}
	\affiliation{\usal}
	
	\begin{abstract}		
		We investigate the chaotic phase of the Bose-Hubbard model [L. Pausch et al, \href{http://dx.doi.org/10.1103/PhysRevLett.126.150601}{Phys. Rev. Lett. \textbf{126}, 150601 (2021)}] in relation to the bosonic embedded random matrix ensemble, which mirrors the dominant few-body nature of many-particle interactions, and hence the Fock space sparsity of quantum many-body systems.
		The energy dependence of the chaotic regime is well described by the bosonic embedded ensemble, which also reproduces the Bose-Hubbard chaotic eigenvector features, quantified by the 
		expectation 
		value and eigenstate-to-eigenstate fluctuations of fractal dimensions. 
		Despite this agreement, in terms of the fractal dimension distribution, these two models  
		depart from each other and from the Gaussian orthogonal ensemble as Hilbert space grows. These results provide further evidence of a way to discriminate among different many-body Hamiltonians in the chaotic regime.  
		
	\end{abstract}
	\maketitle
	
	\section{Introduction}
	Chaotic behavior emerges frequently in both classical and quantum physics \cite{lichtenberg83,ozorio88,ott93,Giannoni89,Izrailev1990,Haake2004}. 
	Classically, the defining property of chaos is the exponential sensitivity of dynamics on the initial conditions \cite{ott93}, giving rise to bundles of wandering trajectories in phase space. Chaos is hence intimately related to ergodicity, which is a feature of just few physical systems and typically difficult to verify \cite{lichtenberg83,ozorio88,ott93}.
	For quantum systems with a strictly chaotic classical limit, 
	statistical properties of the energy spectrum were shown to be 
	in agreement with the (intrinsically ergodic \cite{Pandey1979}) Gaussian ensembles of random matrix theory (RMT) \cite{bgs84,Berry1985,Giannoni89,Guhr1998,Muller2004}. 
	RMT has therefore set the baseline to define chaos in quantum systems: 
	Spectral statistics are commonly used to distinguish chaotic from integrable and localized regimes \cite{delande1986,Stockmann1990,Kolovsky2004,Oganesyan2007,Pal2010,Serbyn2016,Dubertrand2016,Kos2018,Schulz2019a},   
	as are RMT predictions for eigenvector features \cite{Beugeling2018,DeTomasi2020} and other observables \cite{Khaymovich2019}. 
	
	Given that most dynamical systems have a mixed rather than purely chaotic phase space \cite{Geisel1986,Bohigas1993,Buchleitner1995,Ketzmerick2000,Hiller2006,Michailidis2020}, 
	deviations from RMT would be expected \cite{Weidenmuller2009}. Additionally, the standard Gaussian ensembles, described by dense matrices, fail to capture a characteristic feature of many-body Hamiltonians, namely the structural Fock-space sparsity induced by the typical few-body nature of interactions. To account for the latter, further random matrix ensembles, so-called embedded ensembles, have been introduced \cite{Bohigas1971,Bohigas1971a,Mon1975,Benet2003,Kota2014,Chavda2017} with the aim of obtaining a more accurate description of the universal properties of the chaotic regime of many-body quantum systems.
	
	Understanding the conditions for the emergence of ergodicity \cite{Borgonovi2016} or its absence \cite{Abanin2019a}, and the ensuing complex dynamics in many-particle quantum systems is currently a topic of intense research, with ultracold atoms in optical potentials as a prominent experimental platform 
	\cite{Ronzheimer2013a,meinert2014,Preiss2015,Islam2015,Kondov2015a,Schreiber2015b,Choi2016a,Bordia2016,Meinert2016a,Kaufman2016,Bordia2017a,Rispoli2018a,Kohlert2019}, 
	and interacting bosons on regular lattices as a paradigmatic theoretical model, exhibiting 
	signatures of chaos in its spectrum \cite{Kolovsky2004,Kollath2010,Dubertrand2016,Fischer2016a},
	its eigenvectors \cite{Kolovsky2004,Kollath2010,Beugeling2014,Beugeling2015,Beugeling2015c,Beugeling2018} and its time evolution~\cite{Buchleitner2003,Kollath2007,Roux2009,Roux2010,Biroli2010b,Sorg2014}. 
 	
	In a recent publication \cite{Pausch2020}, we have investigated the chaotic and non-chaotic phases of one-dimensional interacting bosons described by the Bose-Hubbard Hamiltonian (BHH). 
	Our study 
	revealed a remarkable agreement with the predictions of the Gaussian orthogonal ensemble (GOE) of RMT, at the level of spectral statistics and the typical value and fluctuations of fractal dimensions for chaotic eigenvectors. Most interestingly, we found that, despite this agreement, both models become ever more distinguishable in terms of the fractal dimension distribution as the dimensionality of Hilbert space grows. This establishes an appealing connection with the current problem of certification of distinctive features of complex quantum systems \cite{Tichy2010,Aolita2015,Walschaers2016,Giordani2018,Zache2020}, and suggests the enticing prospect of having an accessible figure of merit that could unveil the specifics of the underlying many-body Hamiltonian in the chaotic regime, which is otherwise mainly determined by universal features. We nonetheless also contemplated the possibility that such deviation could be accounted for by the more refined embedded ensembles.
	
	The purpose of the present work is to provide a detailed analysis of the chaotic properties of BHH and of the bosonic embedded Gaussian orthogonal random matrix ensemble (EGOE), and to perform a comparison between both models to come to a definite answer to the question posed above.

	The remainder of the manuscript is organized as follows. In sections \ref{sec:model} and \ref{sec:framework}, we present the physical models 
	and the theoretical tools used to characterize their chaotic behaviour. 
	In section \ref{sec:chaosBHH}, we discuss fingerprints of 
	chaos in the BHH spectrum and eigenstates
	as a function of appropriately scaled energy and interaction strength.
	The properties of EGOE and the comparison between both models are first presented in section 
	\ref{sec:EGOEvsBHH}.
	Building upon these results, we investigate, in section \ref{sec:RMT}, the scaling of the chaos markers and of the fractal dimension distributions with Hilbert space dimension, correlating the behaviours of GOE, EGOE and BHH. 
	We summarize our findings and conclude in section \ref{sec:conclusions}.	
	
	\section{Physical Models}
	\label{sec:model}
	
		\subsection{Bose-Hubbard Hamiltonian}
		\label{sec:modelBHH}
		The one-dimensional Bose-Hubbard Hamiltonian \cite{Lewenstein2007,Bloch2008,Cazalilla2011,Krutitsky2016}
		of $N$ bosons on $L$ lattice sites is the sum of a one-particle tunneling observable and a local two-particle interaction, $H=H_\mrm{tun} + H_\mrm{int}$, where 
		\begin{align}
			H_\mrm{tun} &= -J\sum_{j}\left(a^\dagger_j a_{j+1} + a^\dagger_{j+1} a_j\right), \label{eq:H1POint} \\
			H_\mrm{int} &=\frac{U}{2} \sum_{j=1}^L a_j^\dagger a_j^\dagger a_j a_j.
			\label{eq:H2POint}
		\end{align}
		Here, $a_j^\dagger$, $a_j$ are standard bosonic creation and annihilation operators associated with $L$ Wannier orbitals localized at each lattice site. 
		$H_\mrm{tun}$ describes particle tunneling between neighbouring sites with tunneling strength $J$, and  
		$H_\mrm{int}$ provides repulsive on-site interactions with interaction strength $U>0$.
		The sum in the tunneling Hamiltonian runs from $j=1$ to $j=L-1$ for hard-wall boundary conditions (HWBCs) and from $j=1$ to $j=L$ for periodic boundary conditions (PBCs), where we identify $a^{(\dagger)}_{L+1}:=a^{(\dagger)}_1$. 
		For both boundary conditions $H$ exhibits reflection symmetry 
		about the center of the chain, $j\mapsto L+1-j$, 
		and for PBCs it is also invariant under translations, $j\mapsto (j+j')\mod L$. 
		The underlying Hilbert space can therefore be decomposed into a symmetric (even parity, $\pi=+1$) and an antisymmetric (odd parity, $\pi=-1$) subspace with respect to reflection symmetry for HWBCs, while for PBCs the Hilbert space decomposes into subspaces characterized by 
		the total quasimomentum $Q=0,1,\ldots,L-1$ and, in the case of $Q=0$ or $Q=L/2$,
		by further reflection symmetry.
		
		Defining a new set of creation and annihilation operators
		\begin{equation}
		  b_k = \begin{cases}
			\displaystyle
			\sum_{j=1}^L \frac{e^{ -\i\,j \phi(k)}}{\sqrt{L}} a_j, \quad \phi(k)=\frac{2\pi k}{L}, & \text{ for PBCs},\\           
			\displaystyle
			\sum_{j=1}^L \frac{\sqrt{2}\sin\left[j\phi(k)\right]}{\sqrt{L+1}}a_j, \quad \phi(k) = \frac{\pi k}{L+1}, & \text{ for HWBCs},
		        \end{cases}
			\label{eq:bk}
		\end{equation}
		the operators $H_\mrm{tun}$ and $H_\mrm{int}$ can be written as 
		\begin{align}
			H_\mrm{tun} &= -2J\sum_{k=1}^{L} \cos\phi(k)\,b^\dagger_k b_k,
			\label{eq:H1POtun}\\
			H_\mrm{int} &= U \sum_{k,l,m,n=1}^{L} \Delta_{kl}^{mn} b_k^\dagger b_l^\dagger b_m b_n.
			\label{eq:H2POtun}
		\end{align}
		Here, $\Delta_{kl}^{mn}$ accounts for momentum conservation and obeys 
		\begin{equation}
		 \Delta_{kl}^{mn}= \begin{cases}
		                    \dfrac{\delta_{k+l,m+n}}{2L}, & \text{ for PBCs},\\[2mm]
		                    \displaystyle
		                    \sum_{\sigma_i=\pm 1} \frac{\sigma_1\sigma_2\sigma_3}{4(L+1)}\, \delta'_{k+\sigma_1 l, \sigma_2 m + \sigma_3 n}, &\text{ for HWBCs},
		                   \end{cases}
		\end{equation}
		where $\delta_{ij}=1\Leftrightarrow  (i=j)\,\text{mod}\,L$ (PBCs) and 
		$\delta'_{ij}=1\Leftrightarrow {(i=j)\,\text{mod}\,2(L+1)}$ (HWBCs).
		
		Both $H_\mrm{int}$ and $H_\mrm{tun}$, which constitute the respective limits of $H$ for vanishing tunneling ($J=0$) or interaction ($U=0$), are individually integrable, with eigenstates $\ket{\vec{n}} = \ket{n_1,\ldots,n_L}$
		and $\ket{\tilde{\vec{n}}} = \ket{\tilde{n}_1,\ldots,\tilde{n}_{L}}$, respectively, 
		which are uniquely identified by the eigenvalues of 
		the number operators $n_j = a_j^\dagger a_j$ and $\tilde{n}_k = b_k^\dagger b_k$. 
		Natural bases of the BHH are hence symmetrized versions (according to reflection and/or translation symmetry) of the interaction Fock basis $\left\{\ket{\vec{n}} \right\}$ and of the tunneling Fock basis $\left\{\ket{\tilde{\vec{n}}} \right\}$,
		that we refer to as the interaction basis and the tunneling basis, respectively.
		Note that these two bases are conjugated in the following sense:
		The eigenstates of $H$ in the limit $J\to 0$ ($U\to 0$) are localized in the interaction (tunneling) basis but highly delocalized in the conjugated basis.
		
		The non-integrability of $H$ resulting from the 
		combination of interaction and tunneling
		leads to the emergence of a chaotic phase that leaves an imprint in the spectral and eigenvector properties
		\cite{Kolovsky2004,Kollath2010,Dubertrand2016,Fischer2016a,Pausch2020}.
		
		Due to particle number conservation, $[H,N] = 0$ with $N=\sum_j n_j=\sum_k \tilde{n}_k$, we restrict our analysis to subspaces of fixed $N$. 
		In particular, we consider filling factor one, $N=L$, if not stated differently.
		As $N\to\infty$, the BHH reaches its classical limit \cite{Hiller2006,Hiller2009,Dubertrand2016}, whose dynamics is governed only by the scaled energy $H/UN^2$ and the scaled tunneling strength
		\begin{equation}
		 \eta= J/UN .
		 \label{eq:etadef}
		\end{equation}
		As we demonstrated in Ref.~\cite{Pausch2020}, $\eta$ is the parameter that controls the appearance of the chaotic phase in the quantum system at unit density, even for a moderate number of particles.

		We obtain full spectra of the BHH numerically using exact diagonalization, and also calculate eigenstates and eigenenergies around chosen target energies 
		\cite{Pietracaprina2018,petsc-user-ref,slepc}. These calculations are performed for several irreducible symmetry-induced subspaces, for different boundary conditions and in both natural bases. 
		We analyze spectral and eigenvector features for different $N$ as functions of the scaled energy 
		\begin{equation}
		  \eps = (E-E_\mrm{min})/(E_\mrm{max}-E_\mrm{min})
		  \label{eq:Edef}
		\end{equation}
		and $\eta$. Since
		 $H_\mrm{tun} \sim J N = \eta UN^2$
		and $H_\mrm{int} \sim U N^2$ for $N\gg1$ and arbitrary $L$,
		note that $E_\mrm{max}-E_\mrm{min} \sim U N^2$ for fixed $\eta$ and large $N$.
		Hence, $\eps$ scales the energy in the same way as $H/UN^2$, thus effectively providing the relevant energy scale in the classical limit. 
		
		\subsection{Bosonic embedded ensemble}
		The Gaussian Orthogonal Ensemble (GOE), the typical benchmark of quantum chaos in time-reversal invariant systems such as the BHH, consists of fully
		dense matrices, in which each basis state
		is connected to every other.
		In a many-body system, however, many-particle basis states are typically coupled via few-body processes,
		giving rise to a characteristic sparsity in the Hamiltonian's matrix representation. 
		
		To describe such structural constraint of many-body systems, we 
		consider the bosonic two-body GOE embedded ensemble (EGOE), which is defined by the Hamiltonian
		\begin{align}
			H _{\mrm{EGOE}} &= H_1 + \lambda H_2, \label{eq:EGOEdef}\\
			H_1 &= \sum_{i,j=1}^L G^{(1)}_{ij} d_i^\dagger d_j,\\
			H_2 &= \sum_{j\geq i, l\geq k}^L \frac{1}{\sqrt{(1+\delta_{ij})(1+\delta_{kl})}} G^{(2)}_{ij,kl} d_i^\dagger d_j^\dagger d_k d_l, \label{eq:EGOEdef2}
		\end{align}
		where $d_j^\dagger$, $d_j$ are bosonic creation and annihilation operators of $L$ arbitrary orthogonal single-particle states.  
		$G^{(1)}$ and $G^{(2)}$ are independent GOE random matrices in the space of one and two particles, respectively:
		Their matrix elements are independent Gaussian random variables with zero mean and variances 
		$\var(G_{ij}^{(1)}) = 1+\delta_{ij}$, $\var(G_{ij,kl}^{(2)})=1+\delta_{ik}\delta_{jl}$ 
		\cite{Haake2004,Kota2014}. Note that the addition of the single- and two-particle terms leads to 
		correlations in the matrix elements of $H_\mrm{EGOE}$, e.g., its $\binom{N+L-1}{N}$ \cite{Lavenda1991} diagonal entries in the corresponding Fock basis of indistinguishable bosons, 
		$\sum_{i=1}^L G^{(1)}_{ii} \braket{d_i^\dagger d_i} + \sum_{j\geq i}^L G^{(2)}_{ij,ij} \braket{d_i^\dagger d_j^\dagger d_i d_j}/(1+\delta_{ij})$, 
		are defined by just $L$ independent numbers $G_{ii}^{(1)}$ and $L(L+1)/2$ independent numbers $G_{ij,ij}^{(2)}$.
		The parameter $\lambda$ tunes the strength of the interaction part and can be used to induce the emergence of
		chaotic behavior \cite{Kota2014}. 
		
		In order
		to access the same Hilbert space dimensions as for the symmetry-induced subspaces of BHH we demand that $H_\mrm{EGOE}$
		obeys reflection symmetry as BHH. This imposes additional constraints on $G^{(1)}$ and $G^{(2)}$, 
		\begin{align}
			G^{(1)}_{ij} &= G^{(1)}_{(L+1-i)(L+1-j)}, \\
			G^{(2)}_{ij,kl} &= G^{(2)}_{(L+1-i)(L+1-j),(L+1-k)(L+1-l)},
		\end{align}
		that ensure a block diagonal structure
		with respect to parity $\pi$.
		
\begin{figure}
			\includegraphics[width=\linewidth]{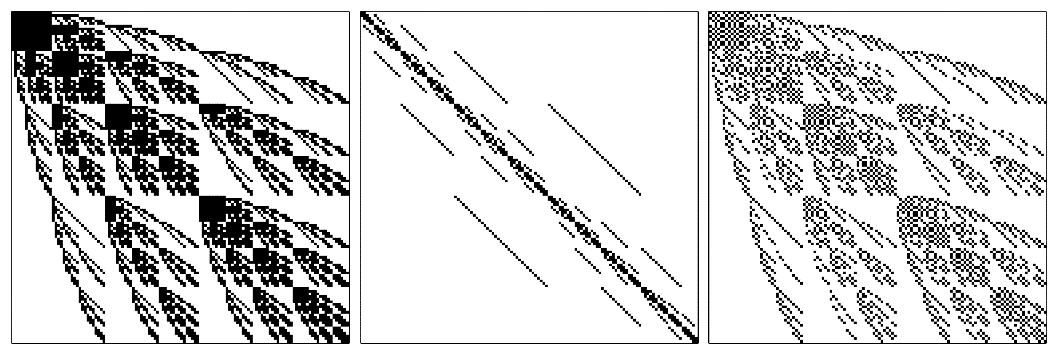}
			\caption{Visualization of the Hamiltonians for a single realization of EGOE [left panel, Eqs.~\eqref{eq:EGOEdef}--\eqref{eq:EGOEdef2}] and for BHH in the interaction basis [center panel, Eqs.~\eqref{eq:H1POint} and \eqref{eq:H2POint}] and in the tunneling basis [right panel, Eqs.~\eqref{eq:bk}--\eqref{eq:H2POtun}] for the full Hilbert space of $N=L=5$ 
			with HWBCs
			(Hilbert space dimension $\NN=126$). 
			Black pixels highlight nonzero matrix elements.}
			\label{fig:matrix-comparison}
		\end{figure}

		By definition, $H_\mrm{EGOE}$ captures the maximal Fock space connectivity allowed by at most two-particle operators. The sparsity of this model (i.e., its zero matrix element structure) is therefore contained in the matrix of the BHH in any basis.
		Figure \ref{fig:matrix-comparison} illustrates the matrix structure of EGOE and of BHH for HWBCs in both natural bases:
		In the interaction basis, different Fock states may be coupled by one-particle nearest-neighbour tunneling only [Eq.~\eqref{eq:H1POint}], and hence the BHH matrix is even sparser than EGOE. In the tunneling basis, however, the two-particle interaction [Eq.~\eqref{eq:H2POtun}] becomes strongly off-diagonal, and the BHH matrix structure closely resembles that of EGOE. 

		Like for the BHH, the spectral and eigenvector features of EGOE matrices will be numerically investigated using exact diagonalization, implementing an ensemble (disorder) average over 100 realizations of $H_\mrm{EGOE}$, unless otherwise stated.

	\section{Quantifiers of Energy Level Statistics and Eigenstate Structure}
	\label{sec:framework}
	
		\subsection{Level spacing ratios}
		A convenient tool to characterize the short-range properties of a quantum system's energy spectrum 
		is given by the level spacing ratios $r_n$, defined as \cite{Oganesyan2007, Pal2010,Atas2013c} 
		\begin{align}
			r_n = \min\left(\frac{s_{n+1}}{s_n}, \frac{s_n}{s_{n+1}}\right) \in[0,1],
		\end{align}
		where $s_n = E_{n+1}-E_n$ is the $n$th level spacing.
		In contradistinction to the distribution of the bare level spacings, the calculation of $r_n$ 
		does not require any---potentially intricate \cite{Gomez2002,Corps2021}---unfolding procedure. 
		For Gaussian random matrix ensembles, an analytic approximation to the probability density of $r$ is known
		\cite{Atas2013c}, which, for the particular case of GOE, yields the mean level spacing ratio $\left< r\right>_{\mrm{GOE}} = 4-2\sqrt{3}\approx 0.536$, in good agreement with 
		 $\left< r\right>_\mathrm{GOE}\approx 0.5307$ obtained from large-scale numerics \cite{Atas2013c}.
	
		\subsection{Generalized fractal dimensions}
		
		Valuable information about the eigenstate structure in Hilbert space is provided by the    
		\emph{generalized fractal dimensions} (GFDs) \cite{Halsey1986,Nakayama2003,Rodriguez2010,Rodriguez2011}. 
		Hilbert-space multifractality indeed seems to be a generic property of many-body Hamiltonians, and has been exploited to characterize the ground-state \cite{Atas2012,Atas2014,Luitz2014,Lindinger2019} as well as the excitation spectrum \cite{Luitz2015,Torres-Herrera2017,Serbyn2017,Pietracaprina2019,Backer2019,Mace2019,Luitz2020} of various systems.
		
		For a state $\ket{\psi} = \sum_{\alpha} \psi_{\alpha}\ket{\alpha}$ in a Hilbert space of dimension $\NN$ spanned by the orthonormal basis $\left\{ \ket{\alpha} \right\}$, 
		the $q$-moments, for $q\in\R^+$, are defined as $R_q = \sum_{\alpha} \left| \psi_{\alpha}\right|^{2 q}$. 
		As $\NN\to\infty$, the $q$-moments of a state that is equally distributed over the full basis scale as $R_q=\NN^{1-q}$, while they become independent of $\NN$ for a state that lives on a fixed finite set of basis elements. 
		For generic states, one may expect the $\NN\to\infty$ scaling  to be of the form $R_q\sim \NN^{(1-q)D_q}$, which defines the generalized fractal dimensions
		$D_q\in[0,1]$.
		We speak of extended ergodic states if $D_q = 1$ for all $q$, while we consider a state as localized if
		$D_q=0$ for $q\geq 1$ \cite{Note1}. 
		A state with $q$-dependent values $0<D_q<1$ is called multifractal in the considered basis. We stress that the GFDs are basis-dependent quantities by definition, and will typically differ among different bases.
		
		In order to study localization
		in finite-dimensional Hilbert spaces, we define finite-size generalized fractal dimensions
		\begin{align}
			\Dq{q} = -\frac{1}{q-1} \frac{\ln R_q}{\ln \NN},
			\label{eq:GFD}
		\end{align}
		which are rescaled versions of the R\'enyi entropies $S_q = -\ln R_q/(q-1)$ \cite{Renyi1960,Stephan2009,Stephan2010,Stephan2011,Misguich2017}.
		The GFDs are then found as the limits $D_q = \lim_{\NN\to \infty} \Dq{q}$.
		We focus on the special values
		\begin{align}
		 \Dq{1}&=\lim_{q\to 1}\Dq{q}=- \sum_{\alpha}\left|\psi_{\alpha}\right|^2\ln\left|\psi_{\alpha}\right|^2/\ln\NN, \\ 
		 \Dq{2}&= -\ln R_2/{\ln \NN}, \\
		 \Dq{\infty} &= \lim_{q\to \infty} \Dq{q}=-\ln\max_\alpha \left|\psi_{\alpha}\right|^2/\ln\NN.
		 \label{eq:Dinf}
		\end{align}
		As $\NN$ grows, these fractal dimensions govern the scaling of $\exp(S)$, where $S$ is the information entropy,
		of the inverse participation ratio  
		$\mathrm{IPR} \equiv R_2 = \sum_{\alpha} \left| \psi_{\alpha}\right|^{4}$  
		\cite{Edwards1972}, and of the maximum intensity of $\ket{\psi}$, respectively. 
		We note that $\Dq{q}$ is a monotonically decreasing function of $q$ \cite{Hentschel1983}, i.e., $\Dq{1}\geq \Dq{2}\geq\Dq{\infty}$ for any state.

	\section{Chaos in the Bose-Hubbard Model}
	\label{sec:chaosBHH}
	We now unveil the chaotic phase of the BHH by investigating the features of its spectrum and eigenstates, as functions of scaled interaction strength $\eta$ [Eq.~\eqref{eq:etadef}] and scaled energy $\eps$ [Eq.~\eqref{eq:Edef}].
	
		\subsection{Localization properties of individual eigenstates}
		
		\begin{figure*}
			\includegraphics[width=.95\linewidth]{\figdir/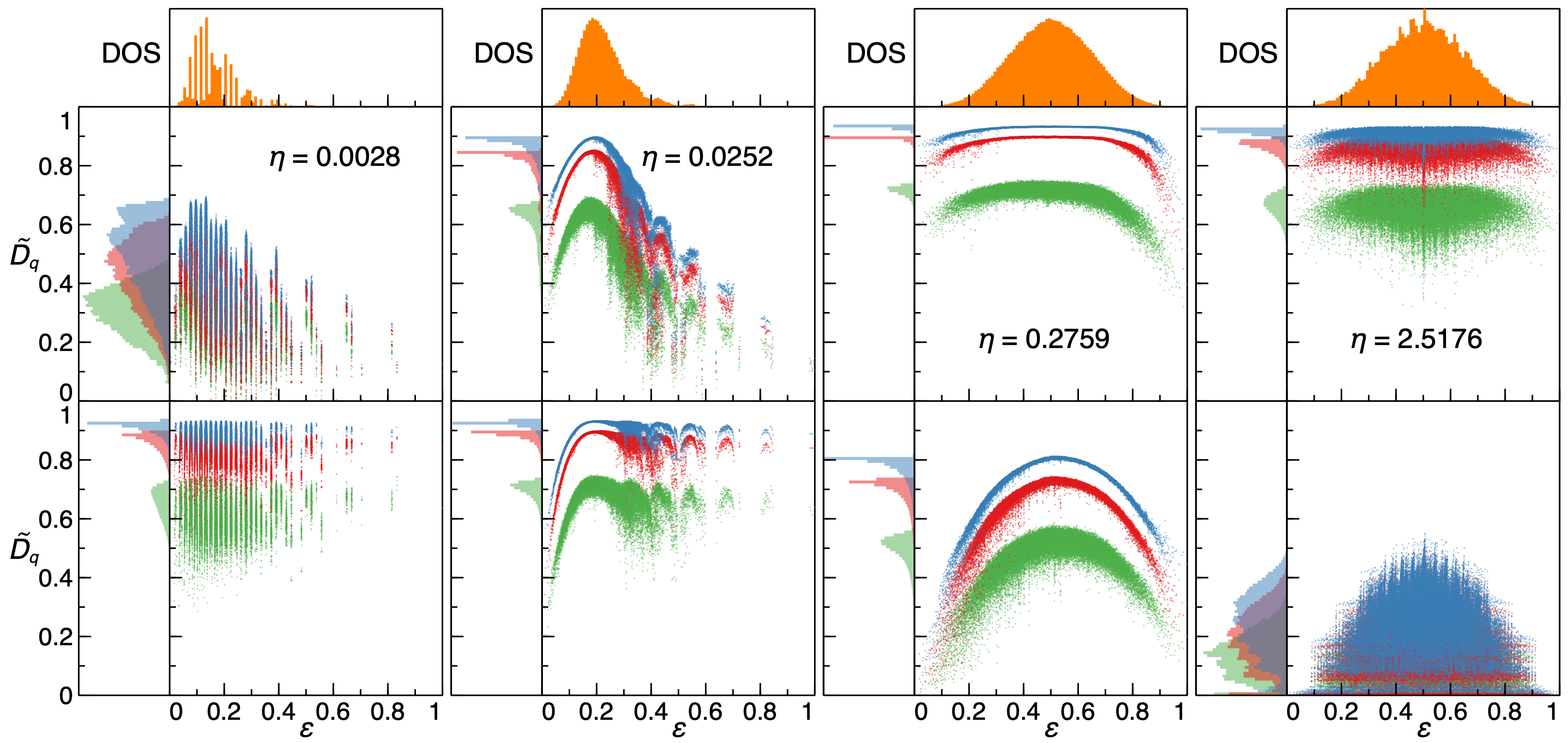}
			\caption{Distribution of the generalized fractal dimensions $\Dq{1}$ (blue), $\Dq{2}$ (red) and $\Dq{\infty}$ (green) of all BHH eigenstates versus scaled energy $\eps=(E-E_\mrm{min})/(E_\mrm{max}-E_\mrm{min})$
				for different values of $\eta= J/UN$ (increasing from left to right) and $N=L=12$ with PBCs, total quasimomentum $Q=0$ and parity $\pi=-1$ ($\NN=55 898$). Upper (lower) main panels show results in the
				interaction (tunneling) basis.
				Plots affixed to the left and top of the main panels display, respectively, histograms of $\Dq{q}$ over all eigenstates, and the density of states (DOS) as a function of $\eps$.}
			\label{fig:scatterplots}
		\end{figure*}

		In Fig.~\ref{fig:scatterplots}, we plot the fractal dimensions $\Dq{q}$ ($q=1,2,\infty$) of all BHH eigenstates, in both natural bases, as functions of
		their scaled eigenenergies $\eps$ 
		for 
		four different values of the parameter $\eta=J/UN$. 
		The subspace considered is that of parity $\pi=-1$ and total quasimomentum $Q=0$, for $N=12$ particles on $L=12$ sites with PBCs ($\NN=55898$).
		Furthermore, we depict the density of states (DOS) and the distributions
		of $\Dq{q}$ over all eigenstates.
		
		The DOS is dominated by discrete peaks
		at small $\eta=0.0028$ (i.e., prevailing interaction), changes to a seemingly Gaussian shape for intermediate $\eta=0.0252$ and $\eta=0.2759$, and reaches a structure featuring several peaks on top of the Gaussian shape at large  $\eta=2.5166$ (i.e., dominating tunneling).
		Upon increasing $\eta$, the DOS shifts from the low energy part of the spectrum
		(with its maximum at $\eps\approx 0.2$) to an almost symmetric shape around $\eps=0.5$.
		
		The behavior of the DOS for extreme values of $\eta$ is inherited from the limiting Hamiltonians. For dominant interaction ($\eta\ll1$), $H$ is close to $H_\mrm{int}$, whose eigenenergies 
		form degenerate manifolds
		at integer multiples of $U$. The multiplicity of the manifolds, and hence the height of the associated DOS peaks is given by the number of permutations of the on-site occupation numbers $\{n_1,\ldots n_L\}$ that give rise to the same eigenenergies. 
		This number is largest, $\sim N!$, when most $n_i$ are different.
		Each particle of such states interacts only with $n_i-1\ll N$ particles, and the corresponding energy lies in the lower half of the spectrum.
		In contrast, high-energy levels correspond to 
		states with $N-n$ particles on a single site, $n\ll N$,
		and their multiplicity is 
		of order $\sim N\binom{N-1}{n}\ll N!$. 
		In the opposite limit ($\eta\gg1$),
		$H$ is close to $H_\mrm{tun}$, whose spectrum is the sum of $N$ identical single-particle spectra. 
		According to the central limit theorem, the DOS should then
		converge to a Gaussian. Deviations from this shape are expected due
		to the discrete nature of the single-particle spectrum for finite $L$.
		
		The fractal dimensions also show a clear development with $\eta$ that is qualitatively the same for all $\Dq{q}$ considered. For \emph{small} $\eta$, the values of $\Dq{q}$
		for states emerging from the same degenerate manifold (for $\eta=0$) are widely spread 
		irrespective of the basis under consideration. 
		In this limit, according to Sec.~\ref{sec:modelBHH}, the eigenstates will tend to be localized in the interaction basis, and hence exhibit lower fractal dimensions than in the tunneling basis, where the eigenstates are more delocalized. 
		Note, however, that full localization (i.e., $\tilde{D}_q\to0$)
		in the interaction basis for small finite $\eta$
		cannot occur,
		since the actual eigenstates as $\eta \to 0$ follow from 
		the diagonalization of the perturbation $H_\mrm{tun}$ on the degenerate manifolds, 
		mixing many Fock states (an ever increasing number with $L$) of the same degenerate manifold.
		
		For \emph{intermediate} $\eta$ values, the $\Dq{q}$ distribution for states that are close in energy becomes very narrow in both bases. 
		This narrowing is enhanced at the bulk of the spectrum---hence anchored to the position of the DOS maximum---, which broadens and shifts to higher energies 
		upon increasing $\eta$ (compare $\eta=0.0252$ and $\eta=0.2759$). 
		In fact, in the interaction basis 
		the distribution of fractal dimensions becomes significantly flat, 
		a behaviour which does not distinctively appear in the tunneling basis.
		Nevertheless, the emergence of a very narrow $\Dq{q}$ distribution in the bulk of the spectrum for intermediate $\eta$
		seems to be a basis-independent feature.
	
		For the \emph{largest} $\eta$ shown, $\eta=2.5166$, the distribution of $\Dq{q}$ broadens again. In the interaction basis
		the fractal dimensions denote a delocalization tendency of the eigenstates (albeit reaching slightly lower $\Dq{q}$ values than for intermediate $\eta=0.2759$), whereas in the tuneling basis
		they tend towards small values without reaching full localization, in agreement with the conjugated nature of both bases discussed earlier and with mixing many degenerate tunneling Fock states in the actual eigenstates \cite{MyNote2}.

		\subsection{Energy-resolved spectral statistics and fractal dimensions}
		
		\begin{figure*}[t]
			\includegraphics[width=.95\linewidth]{\figdir/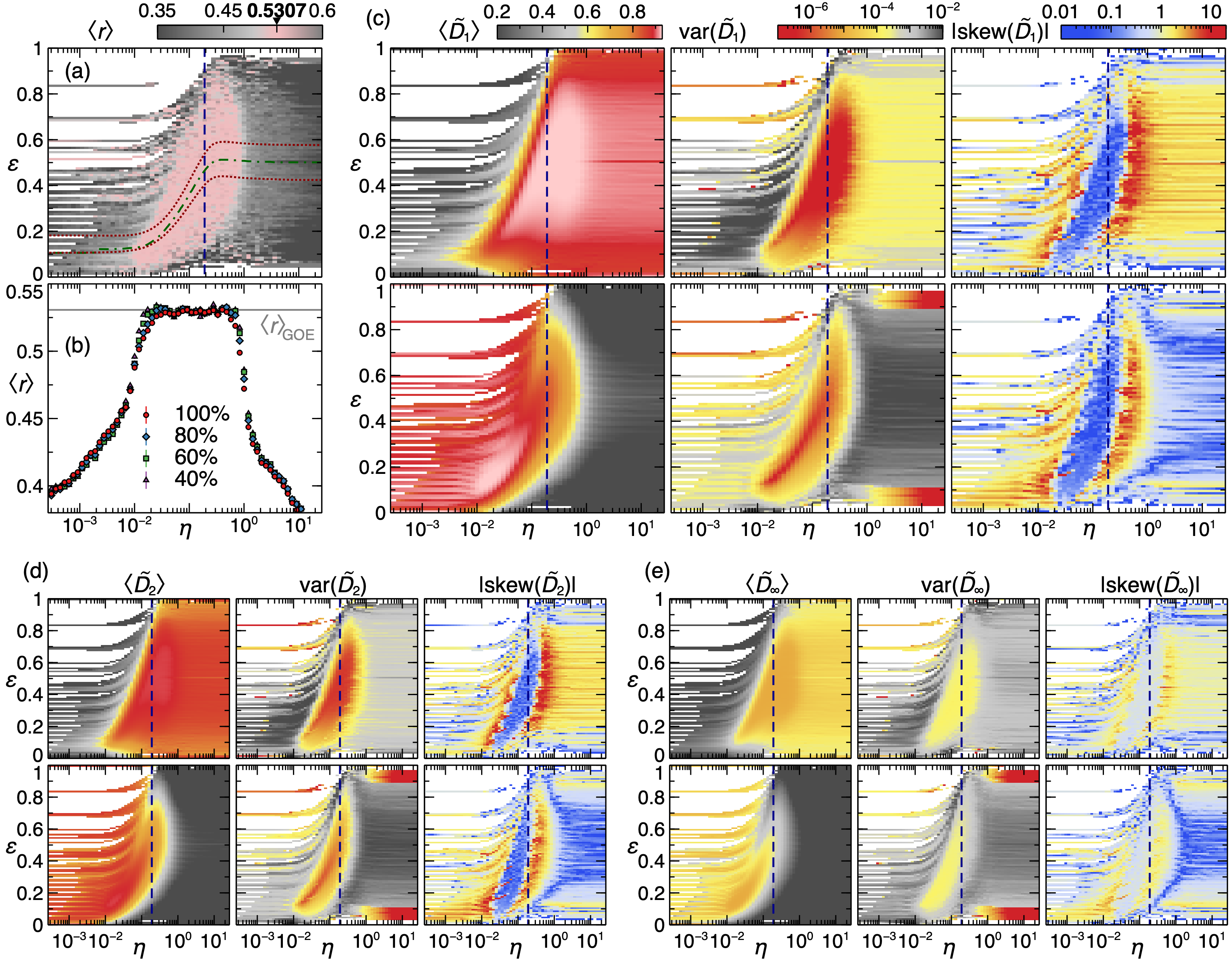}
			\caption{Energy-resolved mean level spacing ratio $\left<r\right>$ (a) and mean, variance and skewness of fractal dimensions $\Dq{1,2,\infty}$ (c)-(e), in the interaction and tunneling bases (upper and lower panels, respectively), as functions of $\eta=J/UN$ and $\eps$ for $N=L=12$ and PBCs [subspace of  $Q=0$ and $\pi=+1$ ($\NN=56822$)]. Panel (b) shows $\left<r\right>$ averaged over the inner $p\%$ of the energy levels ($p=40,60,80,100$) as a function of $\eta$. When not shown, error bars are contained within symbol size.  
			Color scales in (c) apply also to (d) and (e). White areas denote energy bins with too few energy levels to properly define the respective quantity.  Vertical dashed lines highlight the value $\eta=0.1905$ considered in Fig.~\ref{fig:EE_DOS_r_Dqmeanvarskew}. In panel (a), the two red dotted lines demarcate the inner 40\% of the energy levels and the green dash-dotted line indicates the trajectory of the DOS maximum.}
			\label{fig:enresolved}
		\end{figure*}

		We now proceed to investigate systematically the relation between spectral and eigenstate structure changes as functions of $\eta$ and $\eps$.
		To this end, we characterize statistically the distribution of energy levels and fractal dimensions over small energy windows. 

		Figure \ref{fig:enresolved} shows the evolution with $\eps$ and $\eta$ of 
		mean level spacing ratio $\left\langle r\right\rangle$,
		and mean $\aDq{q}$, variance $\vDq{q}$, and absolute value of skewness 
		$\sDq{q} = \big\langle\big(\Dq{q}-\langle\Dq{q}\rangle\big)^3\big\rangle/\vDq{q}^{3/2}$ of fractal dimensions for $q=1,2,\infty$ and 
		$N=L=12$ with PBCs [subspace of $Q=0$ and $\pi = +1$ ($\NN=56822$)].  
		To obtain these statistical quantifiers, the $\eps$ axis is divided into 100 bins of equal width, and the $r$ and $\Dq{q}$ distributions are built from the eigenvalues and eigenstates that fall into each bin at fixed $\eta$. 
		Additionally, Fig.~\ref{fig:enresolved}(b) shows $\left<r\right>$ averaged over the innermost $p\%$ of the energy levels as a function of $\eta$, with $p$ ranging from $100$ to $40$, such that the influence of the edges of the spectrum is progressively reduced. 
		
		The energy-resolved level spacing ratio [Fig.~\ref{fig:enresolved}(a)] 
		signals spectral chaos within a slightly bent oval region for $10^{-2} \lesssim \eta \lesssim 1$ and $0.1\lesssim \eps\lesssim 0.9$, where $\left<r\right>$ approaches the prediction of random-matrix theory, $\left<r\right>_\mathrm{GOE}=0.5307$. 
		The emergence of 
		chaotic spectral statistics depends strongly on the energy.
		In particular, the lower $\eta$ 
		threshold shifts to larger $\eta$ (i.e., stronger tunneling) upon increasing the energy, and the width of the chaotic region (in terms of $\eta$) 
		is reduced towards the edges of the spectrum.  
		The shift of the center of the chaotic phase across the $\eps$-$\eta$ plane 
		can be correlated with the 
		drift of the DOS maximum towards larger $\eps$ upon increasing $\eta$ observed in  
		Fig.~\ref{fig:scatterplots} and indicated by the green dash-dotted line in Fig.~\ref{fig:enresolved} (a).

		The spectrum-averaged level spacing ratio in Fig.~\ref{fig:enresolved}(b)
		also shows 
		agreement with $\left<r\right>_\mathrm{GOE}$ for $10^{-2} \lesssim \eta \lesssim 1$, regardless of whether the average is taken over the full spectrum or just the bulk energy levels. Hence, spectral chaos occurs not only for a wide range of energies, 
		but also for a large majority of the levels.
		The transition into 
		the $\left<r\right>\approx\left<r\right>_\mathrm{GOE}$ plateau
		is rather sharp and becomes even sharper when averaging over a smaller percentage of the inner spectrum. This feature is most notably visible at the plateau's lower $\eta$ edge, and 
		reflects the dependence of the onset of chaos on energy discussed above. 
		Far outside the chaotic region and deep within its center, however, averaging over different percentages of the inner spectrum has only a weak influence on $\left<r\right>$, which shows that, for these regimes, most levels belong to energy regions of similar spectral statistics.

		As observed in Figs.~\ref{fig:enresolved}(c)-(e), the boundary of the spectrally chaotic region correlates with a pronounced increase of the mean fractal dimensions $\aDq{q}$, i.e., with a distinct delocalization tendency of the states in Fock space. In the interaction basis, $\aDq{q}$ grows rapidly from, e.g., $\aDq{1}\approx 0.5$ to $\aDq{1} \gtrsim 0.9$ in a small, $\eps$-dependent $\eta$ range which follows the left boundary of the chaotic region. In the tunneling basis, however, given the conjugated character of the bases, the GFD surge unveils most clearly the upper $\eta$ boundary of the chaotic phase.
		
		Most interestingly, the variation of the finite-size GFDs for close-in-energy eigenstates, captured by $\vDq{q}$, registers a pronounced depression in a parameter domain that correlates remarkably with the spectrally chaotic region. Although this variance shows some slight quantitative differences between the interaction and the tunneling basis in Fig.~\ref{fig:enresolved}(c)-(e), in both cases it exhibits the same qualitative behaviour: It decreases dramatically by several orders of magnitude when entering the chaotic phase. This figure of merit thus appears as a rather sensitive probe which can 
		serve as a basis-independent quantifier of chaos, according to our results. 
		
		The behaviour of mean and variance of $\Dq{q}$ reveals a clear connection between spectral chaos and eigenstate structure:
		When the spectrum indicates chaos in the sense of RMT,
		the involved eigenstates share a similar structure in Fock space, characterized by a marked delocalization tendency in both natural bases.  

		Like the variance, the skewness of the $\Dq{q}$ distribution also exhibits a qualitatively basis independent behaviour: In the $\eps$-$\eta$ parameter space, $\sDq{q}$ undergoes a sharp increase (by at least one order of magnitude) at the boundaries of the chaotic region. This reflects that upon tuning $\eta$ not all eigenstates with similar energies flow into the chaotic region (i.e., toward higher $\Dq{q}$ values) at the same speed, which gives rise to an asymmetric tail in the $\Dq{q}$ distribution, and hence to an enhanced skewness. 
		Within the chaotic region, on the other hand, 
		the fractal dimension distribution is markedly symmetric for $q=1,2$: 
		At the center of the chaotic phase 
		$\big|\sDq{1,2}\big|\lesssim 0.1$
		compared to $\big|\sDq{1,2}\big|>1$ at the boundaries.

		For $q=\infty$, however, 
		the reduction of the skewness in the chaotic region is far less pronounced, and the corresponding distribution there remains distinctively asymmetric as we show in more detail in Sec.~\ref{sec:RMT}.

		As can be seen in Figs.~\ref{fig:enresolved}(c)-(e), the three quantifiers of eigenstate structure considered,  $\aDq{q}$, $\vDq{q}$, and $\sDq{q}$, exhibit the same qualitative behaviour for increasing $q$, albeit the contrast of the chaotic phase is reduced, e.g., the minimum of $\vDq{q}$ becomes less deep for larger $q$ (note that the same color scale applies to all $q$ values). 
		Since $\Dq{\infty}$ is determined by the largest intensity of each state [Eq.~\eqref{eq:Dinf}], already the statistics over one specific intensity in Fock space
		reveals the emergence of chaos.

	\section{EGOE versus Bose-Hubbard}
	\label{sec:EGOEvsBHH}
	
		In the following, we
		investigate how BHH relates to EGOE [Eqs.~\eqref{eq:EGOEdef}-\eqref{eq:EGOEdef2}] within the chaotic phase. This will be done first 
		as a function of energy, for a fixed value of $\eta$ within the BHH's chaotic phase, and subsequently for fixed target energies changing $\eta$.

		\subsection{Comparison of energy dependent features of chaos}
		\label{sec:EGOEvsBHHa}

		\begin{figure*}
			\includegraphics[width=.95\linewidth]{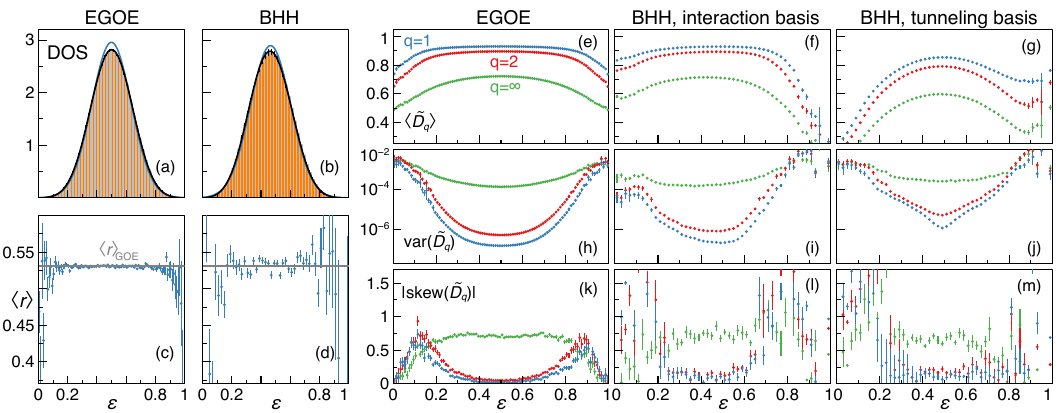}
			\caption{Density of states (a)-(b), mean level spacing ratio $\left<r\right>$ (c)-(d), and mean (e)-(g), variance (h)-(j), and skewness (k)-(m) of $\Dq{1,2,\infty}$
			as functions of energy $\eps$ for the EGOE and the Bose-Hubbard Hamiltonian with HWBCs and $\eta=0.1905$. The subspace considered for both models corresponds to $N=L=10$ and $\pi=-1$ ($\NN=46126$).
			The $\eps$ axis is discretized in 100 (50) equal bins for EGOE (BHH), within which each quantity is calculated, and in the EGOE case further averaged over 100 ensemble realizations (EGOE error bars are solely determined by this average). 
			Solid lines in panels (a) and (b) are fits to a Gaussian distribution (blue) and to a second-order Edgeworth expansion  (black). The horizontal line in (c) and (d) marks the value 
			$\left<r\right>_{\mrm{GOE}}\approx 0.5307$.
			}
			\label{fig:EE_DOS_r_Dqmeanvarskew}
		\end{figure*}

		In order to compare both models, from now on we fix $\lambda=1$ in Eq.~\eqref{eq:EGOEdef} for EGOE, since for this value the chaotic behaviour is fully developed, meaning that the studied figures of merit do not change upon further increase in $\lambda$, for any value of the energy $\eps$, see appendix \ref{sec:EGOEvsLambda}. For BHH the scaled tunneling strength is set to 
		$\eta=0.1905$, which is indicated by vertical dashed lines in Fig.~\ref{fig:enresolved} and is deep within the chaotic phase. 
		Figure \ref{fig:EE_DOS_r_Dqmeanvarskew} provides an overview of the spectral and eigenstate structural features, quantified in terms of the finite-size GFDs, of the two Hamiltonians as functions of $\eps$, for the $\pi=-1$ Fock subspace of the $N=L=10$ system ($\NN=46126$). Mean level spacing ratios and moments of $\Dq{q}$ distributions are obtained after discretizing the energy axis, as explained above, and for EGOE they follow from an additional average over 100 ensemble realizations. 
		It must be stressed that the ensemble-averaged EGOE spectral and eigenstate properties are symmetric around $\eps=0.5$ by definition [Eqs.~\eqref{eq:EGOEdef}--\eqref{eq:EGOEdef2}], while such symmetry does in general not hold for the BHH (it emerges only for sufficiently large $\eta$). For the $\eta$ value considered, the BHH's DOS starts approaching such symmetric behaviour, which enables the model comparison for the same $\eps$. Nonetheless, the BHH features may naturally exhibit an asymmetry at the spectrum's tails which cannot be quantitatively captured by EGOE. 
		
		In contrast to GOE, whose DOS is dictated by the semi-circle law \cite{Haake2004}, 
		the EGOE and the BHH energy level distributions [Figs.~\ref{fig:EE_DOS_r_Dqmeanvarskew}(a) and (b), respectively] are found to exhibit a comparable 
		Gaussian-like shape, centered around $\eps=0.5$ ($\eps=0.45$) for EGOE (BHH).
		Close inspection reveals clear deviations from Gaussianity in the DOS that are in both cases accurately described by an Edgeworth expansion up to second order [cf.~blue versus black lines in panels (a) and (b)] \cite{Kota2014}, in which a Gaussian is multiplied by a
		polynomial determined by the third and fourth central moments. Such expansion has also been found to apply to fermionic embedded ensembles \cite{Kota2014}.
		
		The chaotic phase in both models is revealed by the agreement of the mean level spacing ratio with the GOE value in the range $0.2\lesssim \eps\lesssim0.8$, flanked by strong fluctuations of $\left\langle r\right\rangle$ at the spectral edges [Fig.~\ref{fig:EE_DOS_r_Dqmeanvarskew}(c) and (d)]. Note that the overall smoother behavior of $\left\langle r\right\rangle$ for EGOE is a consequence of the additional ensemble average. 
		The analysis of $\left\langle r\right\rangle$ shows that EGOE captures the absence of chaos at the spectral edges in qualitative agreement with the BHH. This is in great contrast to GOE, whose spectral (as well as eigenvector) properties do not carry any dependence on energy \cite{Haake2004,Brody1981,Guhr1998}, and suggests that sparse Fock space connectivity must be responsible for the features observed at the spectrum's tails. 
		
		Within the spectrally chaotic region, 
		for the three cases analyzed (EGOE and BHH in both natural bases),
		$\aDq{q}$ rise to their maximum values, and simultaneously $\vDq{q}$ achieve their minima after dropping by several orders of magnitude, as shown in Fig.~\ref{fig:EE_DOS_r_Dqmeanvarskew}(e)-(j).
		The maxima of $\aDq{q}$ develop into distinct plateaux for EGOE and the BHH in the interaction basis, which show a very good quantitative agreement [cf.~(e) and (f)], in contrast to the results in the tunneling basis where no clear plateau appears [(g)] for the same Hilbert space dimension. 
		This behaviour is accompanied by a levelling of $\vDq{q}$ around very low values that agree best between EGOE and BHH in the interaction basis. 
		
		As discussed in the previous section, the skewness of the finite-size GFDs was particularly sensitive to the boundaries of the chaotic phase. 
		This sensitivity is manifestly shown as a function of the energy in panels (k)-(m). 
		In both models, the absolute skewness for $q=1$ and $q=2$ is suppressed within the chaotic region, with comparable values in all cases, and maximal at its borders, which for BHH also correlate with a clear increase in the uncertainties. 
		In contrast to $q=1,2$, the skewness for $q=\infty$ is not suppressed within the chaotic region but rather acquires a common value $\left|\sDq{\infty}\right|\approx 0.7$ in all three cases, indicating the persistent asymmetry of the $\Dq{\infty}$ distribution.  
		
		Although EGOE captures the major qualitative features of the finite-size GFDs' behaviour in the BHH, it must be noticed that the agreement is manifestly better with the analysis in the interaction basis, despite the fact that the EGOE Fock space connectivity is far closer to the one of the BHH in the tunneling basis [Fig.~\ref{fig:matrix-comparison}]. 
		In BHH, the correlations among the nonzero connectivity amplitudes in Hilbert space are much stronger than for EGOE, since BHH is defined by the  parameters $J$ and $U$ only, while EGOE depends on all independent entries of the matrices $G^{(1)}$ and $G^{(2)}$. 
		Our findings thus show that these correlations also play a significant role.		
		
		\subsection{Chaos around specific target energies}

		\begin{figure*}
			\includegraphics[width=.95\linewidth]{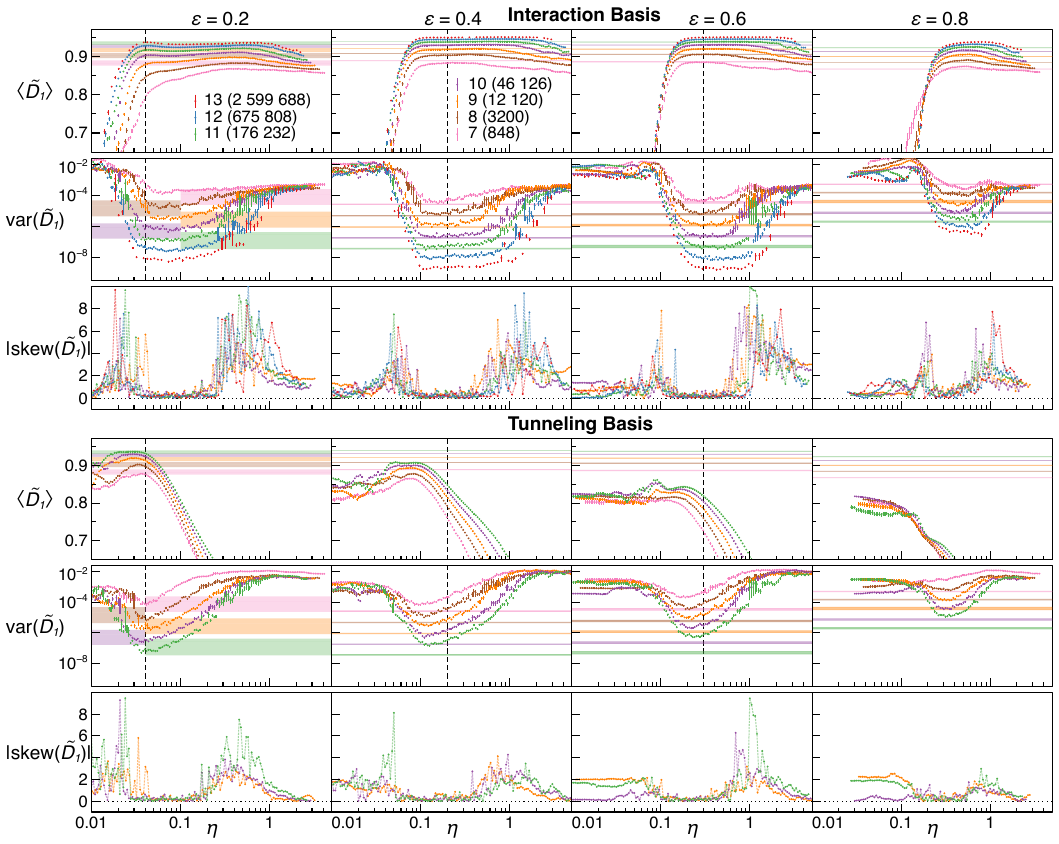}
			\caption{Mean, variance and absolute skewness of $\Dq{1}$ in the interaction basis (upper panels) and the tunneling basis (lower panels) as a function of $\eta=J/UN$ for 100 eigenstates closest to energies $\eps=0.2,0.4,0.6,0.8$ (left to right) for HWBCs with parity $\pi=-1$ and system sizes ranging from $N=L=7$ ($\NN=848$) to $N=L=13$ ($\NN=2\,599\,688$).
			Shaded areas indicate EGOE values (for numerically reachable $N\leq 11$) from the corresponding energy intervals as described in the main text and appendix~\ref{sec:epsEGOE}: $\left(\eps=0.2, \eta\in[0.03,0.15]\right)\to \eps_\text{EGOE}\in[0.25,0.5]$, $\left(\eps=0.4, \eta\in[0.08,1]\right)\to \eps_\text{EGOE}\in[0.4,0.5]$, $\left(\eps=0.6, \eta\in[0.15,1]\right)\to \eps_\text{EGOE}\in[0.6,0.65]$, $\left(\eps=0.8, \eta\in[0.25,0.9]\right)\to \eps_\text{EGOE}=0.8$.
			Error bars are contained within symbol size if not explicitly shown (not included for the skewness to ease visualization). Vertical dashed lines mark the $\eta$ values considered in Fig.~\ref{fig:DqVarRMT}.}
			\label{fig:Dq_vs_eta_allEps}
		\end{figure*}
		
		Figure \ref{fig:Dq_vs_eta_allEps} shows the evolution of $\aDq{1}$, $\vDq{1}$ and $\big|\sDq{1}\big|$ for the BHH as functions of $\eta$ around specific target energies $\eps$ ranging from $0.2$ to $0.8$ and system sizes $N=L\in[7,13]$. These statistical quantifiers are obtained from the 100 eigenstates closest to the considered energy. 
		
		In the interaction basis, for all system sizes and $\eps$ values, $\aDq{1}$ 
		registers a sharp surge upon increasing $\eta$ that tends to develop into a plateau as $N$ grows. 
		For each system size, the onset and width of the plateau depend on energy, in accordance with the chaotic phase qualitatively identified in Fig.~\ref{fig:enresolved}.
		The emergence of the plateau is mirrored by a synchronous drop in the variance of $\Dq{1}$ by several orders of magnitude, towards a flat minimum which becomes ever lower with increasing $L$. 
		This behaviour is also accompanied by a distinctive suppression of the skewness of the $\Dq{1}$ distribution. As discussed above, the emergence of chaos can be most clearly observed in the  behaviour of $\vDq{1}$. The chaotic phase is more prominent [lower $\vDq{1}$ and wider plateaux] for $\eps\leq0.6$ for which there is always a range of $\eta$ values 
		where the corresponding $\eps$ falls within the bulk of the spectrum, as can be observed in Fig.~\ref{fig:enresolved}(a) from the evolution of the inner 40\% of energy levels as a function of $\eta$.
		Energy $\eps=0.8$ can be considered to belong to the tail of the spectrum for all $\eta$, and nonetheless chaos leaves a distinct imprint in the variance of $\Dq{1}$. 
		
		For increasing system size, the left edges of the plateaux
		show a tendency to converge to an $L$ independent $\eta$ value (although depending on $\eps$), which confirms that the semiclassically inspired \cite{Hiller2006,Hiller2009,Dubertrand2016} parameter $\eta=J/UN$ 
		indeed controls the transition into the chaotic phase. 
		Remarkably, upon increasing $L$ the plateaux' right termination point shifts towards larger $\eta$ \cite{SMletter}, indicating that the chaotic region might extend to arbitrarily large $\eta$ (arbitrarily small interactions) in the thermodynamic limit (with a discontinuity at the integrable point $\eta=\infty$). This latter observation agrees 
		with results for fermions \cite{Neuenhahn2012}, spin systems and hard-core bosons \cite{Santos2010a,Torres-Herrera2014}, where 
		the critical perturbation strength to drive an integrable system into chaos was found to decay linearly with system size. 
		
		In order to compare the BHH chaotic behaviour at fixed $\eps$ against EGOE, we must take into account the $\eta$-dependent shift of the spectrum bulk for BHH. In other words, the identification of energy bulk and tails for both models is in general not a one to one correspondence in $\eps$: The EGOE maximum DOS is always located at $\eps=0.5$, while this is not the case for BHH, even in its chaotic phase. We will take the position of the DOS maximum as the bulk centre. Therefore, once we have located the chaotic regime in $\eta$ at fixed $\eps$ for BHH [mainly identified by the behaviour of $\vDq{1}$], we analyze the distance of $\eps$ from the bulk centre for such $\eta$ range, and find the corresponding energies in EGOE (which we refer to as $\eps_\textrm{EGOE}$ in the caption of Fig.~\ref{fig:Dq_vs_eta_allEps}) which lie at the same distance from its bulk centre (see appendix \ref{sec:EGOEvsLambda}). In this way we compare on an equal footing for both models the influence of energy deviation from the bulk centre on the considered markers of chaos. 
		
		As seen in Fig.~\ref{fig:Dq_vs_eta_allEps}, the GFD figures of merit within the BHH chaotic phase for all $\eps$ considered agree remarkably well with the EGOE values for the corresponding system size. This shows that the change of the BHH chaotic phase across the energy spectrum is correctly captured by EGOE, once the above subtlety in the energy comparison is taken into account. 
		
		In contrast to the interaction basis, for the system sizes studied, the formation of $\aDq{1}$ plateaux in the tunneling basis is not so clear, and arguably only for $\eps=0.2$ can they be seen in agreement with the EGOE value. Despite this, and most interestingly, the development of the chaotic phase for increasing $L$ is unmistakably exposed for all $\eps$ by the behaviour of the variance and the skewness of $\Dq{1}$, although agreement with EGOE is not observed for $\eps>0.2$. This reflects the natural basis dependence of chaotic eigenvector features, which might also translate into different scaling behaviours with system size. 
		
	\section{Scaling with Hilbert Space Dimension}
	\label{sec:RMT}
	
	We now analyse the scaling behaviour of the GFDs for BHH and EGOE with Hilbert space dimension $\NN$, 
	in comparison to analytical results for GOE, and determine whether or not the
	models approach each other at the level of the full GFD distribution within the chaotic regime.
	
	\subsection{Scaling of mean and variance}
	\label{sec:ScalingMeanVar}
	
	\begin{figure*}

		\includegraphics[width=.95\linewidth]{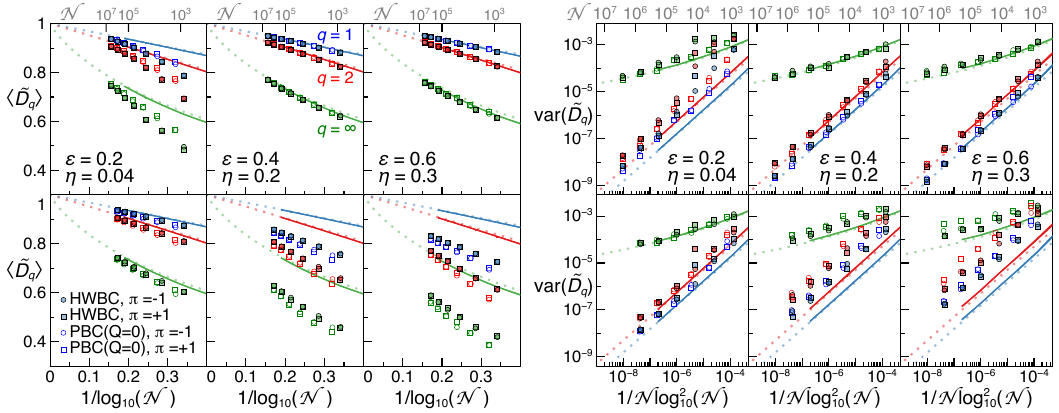}
		\caption{Dependence of $\aDq{q}$
			(left) and $\vDq{q}$
			(right), $q=1,2,\infty$, on Hilbert space dimension $\NN$ in the interaction (upper panels) and in the tunneling basis (lower panels), at three different $(\eps,\eta)$, for four different subspaces distinguished by symbols as indicated. Each data point is computed from 100 eigenstates closest to $\eps$ at the given $\eta$.
			Dotted lines indicate analytical results for GOE from Eqs.~\eqref{eq:meanD1GOE}-\eqref{eq:kthMomentDinf}, while solid lines 
			indicate EGOE data
			for the following energies $\eps_\text{EGOE}$, which are obtained as described in appendix~\ref{sec:epsEGOE} and Fig.~\ref{fig:appendix_epsEGOE}: $\left(\eps=0.2,\eta=0.04\right)\to \eps_\text{EGOE}=0.5$, $\left(\eps=0.4,\eta=0.2\right)\to \eps_\text{EGOE}=0.45$, $\left(\eps=0.6,\eta=0.3\right)\to \eps_\text{EGOE}=0.6$. 
			Whenever not shown, error bars are contained within symbol size.}
		\label{fig:DqVarRMT}
	\end{figure*}
	
	Figure \ref{fig:DqVarRMT} shows the dependence on $\NN$ of $\aDq{q}$ and $\vDq{q}$ of both models
	for three pairs of $(\eps,\eta)$. We restrict to $\eps\leq0.6$ for which, as discussed above, the BHH chaotic phase is more prominent. 
	The corresponding $\eta$ is fixed to ensure that
	$\vDq{1}$ for BHH reaches minimal values either in the tunneling [$(0.2,0.04)$] or in the interaction basis [$(0.4,0.2)$ and $(0.6,0.3)$] (cf.~dashed vertical lines in Fig.~\ref{fig:Dq_vs_eta_allEps}).
	Results for EGOE are obtained from 100 eigenstates closest to the corresponding energy targets $\eps_\textrm{EGOE}$, as described above, and averaged over 100 ensemble realizations.
	We further compare the finite-size expectation values of $\aDq{q}$ and $\vDq{q}$ against the following analytical results for GOE \cite{Pausch2020,SMletter,D2note}: 
	\begin{align}
		\aDq{1}_\mrm{GOE} &= \frac{h_{\NN/2}-2+\ln 4}{\ln \NN}, \label{eq:meanD1GOE}\\
		\tilde{\mathcal{D}}_2^\mrm{GOE} &\equiv  -\log_\NN \langle R_2\rangle_\mrm{GOE} = \frac{\ln(\NN+2)-\ln(3)}{\ln \NN}, \label{eq:varD1GOE}\\
		\vDq{1}_\mrm{GOE} &= \frac{(3\pi^2- 24)(\NN+2) - 8}{2(\NN+2)^2 \ln ^2\NN}-\frac{\psi^{(1)}(2+\NN/2)}{\ln ^2\NN}, \\
		\var\left(\tilde{\mathcal{D}}_2^\mrm{GOE}\right)
		&= \frac{8(\NN-1)}{3(\NN+4)(\NN+6)\ln^2(\NN)}, \label{eq:varD2GOE}
	\end{align}
	where $h_{n}=\sum_{k=1}^n k^{-1}=\int_0^1 (1-x^n)/(1-x) \; \mathrm{d}x$ is the harmonic number, $\psi^{(1)}$ is the first derivative of the digamma function $\psi$ 
	\cite{NIST:DLMF}, and the corresponding quantities for $q=\infty$ are obtained using 
	\begin{align}
		\big\langle\Dq{\infty}^k\big\rangle_\mrm{GOE} \simeq&\frac{k (-1)^{k+1}}{\left(\ln \NN\right)^k} \int_{0}^{1} \mathrm{d} s\;   \frac{\ln(s)^{k-1}}{s}\operatorname{Erf}\left(\sqrt{\frac{\NN s}{2}}\right)^\NN, \label{eq:kthMomentDinf}
	\end{align}
	for integer $k$. The asymptotic behaviour that ensues is
	\begin{align}
		1-\aDq{1}_\mrm{GOE}&\sim \frac{1}{\ln \NN}, 
		\label{eq:D1GOEasym}\\
		\vDq{1}_\mrm{GOE}&\sim \frac{1}{\NN \ln^2 \NN},
	\end{align}
	which also applies to $\tilde{\mathcal{D}}_2^\mrm{GOE}$, and 
	\begin{align}
		1-\aDq{\infty}_\mrm{GOE}&\sim\frac{\ln(\ln\NN)}{\ln \NN}, \\
		\vDq{\infty}_\mrm{GOE}&\sim \ln^{-4}\NN. \label{eq:vDinfGOEasym}
	\end{align}
	
	As observed in Fig.~\ref{fig:DqVarRMT}, $\aDq{q}$ and $\vDq{q}$ are very close for both random ensembles
	and almost coincide for the largest accessible $\NN$. We can then conclude that the EGOE values essentially follow the GOE behavior, in particular that the EGOE dominant finite-size corrections are given by Eqs.~\eqref{eq:D1GOEasym}-\eqref{eq:vDinfGOEasym}.
	
	In the tunneling (for $\eps=0.2$) and interaction basis (for $\eps=0.4$, $0.6$), 
	$\aDq{q}$ and $\vDq{q}$ approach rather quickly, 
	for all $q$ and all symmetry-induced subspaces, 
	the corresponding (E)GOE values when increasing
	$\NN$. Such convergence can also be inferred in the interaction basis at $\eps=0.2$, albeit much slower. 
	For these cases, the numerical evidence indicates that 
	$\aDq{q}$ [$\vDq{q}$] takes the ergodic value
	1 [0] in the thermodynamic limit, and does so according to Eqs.~\eqref{eq:D1GOEasym}-\eqref{eq:vDinfGOEasym}.

	In contrast, in the tunneling basis at $\eps=0.4$ and $0.6$, both statistical quantifiers show pronounced quantitative deviations from the (E)GOE trajectories (as also observed in Fig.~\ref{fig:Dq_vs_eta_allEps}), to the point that a linear extrapolation in $1/\ln \NN$ of the available $\aDq{q}$ data actually yields $\lim_{\NN\to \infty}\aDq{q}<1$, indicating that the dominant finite size correction of BHH in this case must have a different dependence on $\NN$ than (E)GOE.
	This observation clearly manifests that the development of the chaotic phase in the eigenvector features and the scaling trajectory followed towards ergodicity
	depend strongly on the basis considered.
	
	\subsection{Comparison of the full distributions}
	
	\begin{figure*}
		\includegraphics[scale=0.95]{\figdir/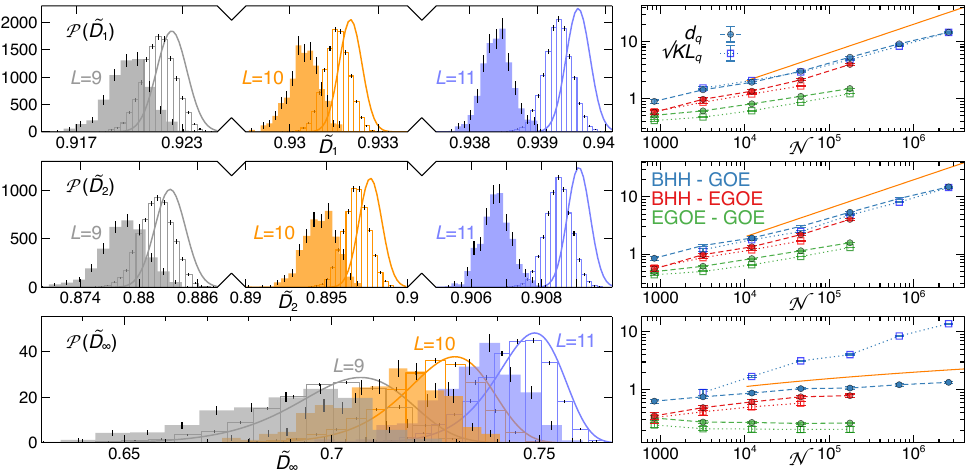}
		\caption{Scaling of GFD distributions $\mathcal{P}(\Dq{q})$ with Hilbert space dimension $\NN$ for $q=1$ (top), $q=2$ (centre), and $q=\infty$ (bottom). Left panels show the distributions for BHH [interaction basis, HWBCs, $\pi=-1$] (filled histograms), EGOE (outlined histograms), and GOE (solid lines) for $N=L=9$ ($\NN=12120$), $N=L=10$ ($\NN=46126$) and $N=L=11$ ($\NN=176232$).
		The BHH histograms are obtained from five $\eta \in [0.25,0.38]$ and 100 eigenstates around $\eps=0.5$ for each $\eta$.
		The EGOE histograms result from 100 realizations and, per realization, 100 eigenstates around $\eps_\text{EGOE}=0.5$.
		The $\Dq{1,2}$ distributions for $L=9$ ($L=10$) were multiplied by 
		4 (2) for better visualization. 
		The scaling of the pairwise distance between distributions is shown in the right panels, and quantified via $d_q$ [Eq.~\eqref{eq:dq}] and the square root of the Kullback-Leibler divergence [Eq.~\eqref{eq:Kullback-Leibler}].
		Solid lines denote the scalings $d_{1,2}\sim \sqrt{\NN}$ and $d_\infty\sim\ln(\ln\NN)\ln\NN$ that would result if the coefficient of the dominant finite-size correction to $\aDq{q}$ for GOE and the other two models were to be different.}
		\label{fig:DqDistrRMT}
	\end{figure*}
	
	In Ref.~\cite{Pausch2020} we have shown that, while $\aDq{q}$ and $\vDq{q}$ of BHH eigenstates converge towards GOE, both models depart from each other at the level of the full GFD distributions with increasing $\NN$. The corresponding analysis for BHH and EGOE will determine whether
	the two-body Hamiltonian structure is what sets BHH apart from GOE.
	
	We compute the distributions $\mathcal{P}(\Dq{1,2,\infty})$ for BHH in the interaction basis (HWBCs, $\pi=-1$) at $\eps=0.5$, where the agreement with EGOE and GOE is best (cf. Fig.~\ref{fig:EE_DOS_r_Dqmeanvarskew}),
	from a total of 500 eigenstates obtained within the chaotic regime for $\eta\in[0.25,0.38]$.
	Similarly, we consider 100 EGOE realizations, per realization using the 100 eigenstates closest to $\eps_\text{EGOE}=0.5$.
	For GOE, we have demonstrated in Ref.~\cite{SMletter} that $\P_\mrm{GOE}(\Dq{1,2})$ are very well described by Gaussian distributions, with mean and variance given by Eqs.~\eqref{eq:meanD1GOE}-\eqref{eq:varD2GOE}, and that $\P_\mrm{GOE}(\Dq{\infty})$ can be taken to be 
	\begin{align}
		\P(\Dq{\infty}) = \frac{\NN^{3/2}}{\sqrt{2\pi t}} e^{-t \NN/2} \left[\operatorname{Erf}(\sqrt{t \NN/2})\right]^{\NN-1} t \ln\NN,
	\end{align}
	with $t=\NN^{-\Dq{\infty}}$.

	As the left panels in Fig.~\ref{fig:DqDistrRMT} reveal,
	the GFD distributions of BHH and EGOE drift apart as $\NN$ grows.
	In the right panels,
	we quantify the distance between the distributions by the relative difference of the means, 
	\begin{equation}
	d_q = \left(\langle\Dq{q}\rangle_\mrm{(E)GOE}-\langle\Dq{q}\rangle_\mrm{BHH}\right)/\sqrt{\vDq{q}_\mrm{BHH}},
	\label{eq:dq}
	\end{equation}
	and by the square root of the Kullback-Leibler divergence (information entropy) \cite{KullbackLeibler,InformationTheory}
	\begin{align}
			KL_q(\P,\mathcal{Q}) = \int_{0}^1 \P\left(\Dq{q}\right) \ln\left( \frac{\P(\Dq{q})}{\mathcal{Q}(\Dq{q})}\right) \; \mrm{d}\Dq{q}.
			\label{eq:Kullback-Leibler}
	\end{align}
	Both distance measures increase with Hilbert space dimension for BHH versus either GOE or EGOE for all $q$ considered, although the growth is less pronounced for $q=\infty$. 
	Note that the EGOE distribution is closer to BHH than GOE, and hence the distance $d_q(\text{EGOE-BHH})$ is always bounded from above by $d_q(\text{GOE-BHH})$.
	For $\NN\geq 10^6$ the mean $\Dq{1,2}$ of BHH differs from GOE already by $10\sigma$, and, by extrapolation, the distance from EGOE, although smaller, will be of comparable magnitude. 
	
	We therefore conclude that the statistical departure between BHH and GOE in terms of the GFD distributions, cannot be solely explained by the two-body nature of BHH and the consequent 
	Fock space sparsity, since these features are also present in EGOE. 
	The observed deviation must then be ascribed to the very specific nature of the BHH on-site interactions and nearest-neighbour tunneling, and the existing correlations among the Hamiltonian matrix elements. 
	
	Interestingly, an increasing trend of $d_q$ and $KL_q$ for $q=1,2$ is also observed
	for EGOE versus GOE, whereas for $q=\infty$ the GFD distributions of both models appear to approach each other in the accessible $\NN$ range. 
	Hence, despite the quantitative agreement exhibited by $\aDq{q}$ and $\vDq{q}$, random two-body Hamiltonians become more distinguishable from full GOE in terms of $\mathcal{P}(\Dq{1,2})$ when approaching 
	the thermodynamic limit. 

	The comparison of the numerical results for BHH (with respect to the interaction basis, at the center of the spectrum) and EGOE with the analytical GOE predictions will help us understand the origin of 
	the departure of $\mathcal{P}(\Dq{1,2})$ in terms of the finite-size corrections to $\aDq{q}$. 
	As discussed in section \ref{sec:ScalingMeanVar}, the numerical results suggest that the functional dependence of the dominant correction for all models is given by Eqs.~\eqref{eq:D1GOEasym}-\eqref{eq:vDinfGOEasym}. 
	Consequently, since $\sqrt{\vDq{1,2}_\mrm{BHH}}\sim 1/\sqrt{\NN}\ln\NN$
	and
	$1-\aDq{1,2}\sim 1/\ln\NN$ for all models, we can single out three main possibilities leading to three different scaling behaviours of $d_q$: 
	(i) the coefficient of the dominant correction differs between GOE and the other models, which leads to $d_{1,2}\sim \sqrt{\NN}$ [indicated by solid lines in the right panels of Fig.~\ref{fig:DqDistrRMT}], 
	(ii) the dominant correction is exactly the same in all models, and BHH and EGOE carry subleading terms in $\aDq{1,2}$ decaying slower than $1/\sqrt{\NN}\ln\NN$, which entails an increasing behaviour of  $d_{1,2}$ slower than $\sqrt{\NN}$, and (iii) the dominant correction is exactly the same in all models, and the second term in BHH and EGOE decays faster than $1/\sqrt{\NN}\ln\NN$, 
	which would imply a decreasing behaviour of $d_{1,2}$, and hence an approach of the distributions.
	
	The tendencies shown in the right panels of Fig.~\ref{fig:DqDistrRMT}  suggest that case (ii) holds. 
	This implies that asymptotically $1-\aDq{1,2}= c_1/\ln\NN$, with the same $c_1$ for the three models. Therefore, the increasing distinguishability of the distributions cannot be attributed to the so-called `weak ergodicity' \cite{Backer2019}.
	In fact, as described above, a common $c_1$ is still compatible with either a departure or a convergence of the distributions, a behaviour that is determined by the functional dependence of the second finite $\NN$ correction.
	
	A similar argumentation applies to $q=\infty$: Were the coefficient of the dominant correction in $\aDq{\infty}$ to differ between GOE and the other models, one should observe $d_\infty\sim \ln(\ln\NN)\ln\NN$. In this case however, given the slow logarithmic increase, it is more challenging to draw a definite conclusion about BHH. We also note that the second and third GOE corrections decay slower than $\sqrt{\vDq{\infty}}\sim\ln^{-2}\NN$ \cite{Pausch2020}, which implies that an increasing behaviour of $d_\infty$ is still compatible with BHH having the first three finite-size corrections in $\aDq{\infty}$ dictated by GOE. For EGOE on the other hand, the decreasing behaviour of $d_\infty$ indicates that both random ensembles do share at least all correction terms that decay slower than $\ln^{-2}\NN$.
	
	\section{Conclusions}
	\label{sec:conclusions}
	We have scrutinized the properties of the chaotic phase of the Bose-Hubbard Hamiltonian (BHH) via spectral and eigenvector features in relation to the bosonic embedded Gaussian orthogonal random-matrix ensemble (EGOE). 
	The emergence of chaos at the level of individual eigenstates as well as statistically over energy intervals for both natural bases of BHH has been analyzed in detail. 
	We have confirmed that the fluctuations of the fractal dimensions for close-in-energy eigenstates are an efficient, and qualitatively basis independent, probe of chaos. Furthermore, the boundary of the chaotic region is signalled by a strongly skewed distribution of the fractal dimensions for close-in-energy eigenstates.
	
	The analysis of the chaotic phase for EGOE in terms of spectral statistics and eigenstate fractal dimensions reveals the same dependence on energy as for BHH, capturing in particular the disappearance of chaos at the spectral edges, in contrast to GOE.
	We therefore attribute this feature to the sparse Fock-space connectivity of the Hamiltonian. 
	EGOE describes remarkably well the typical value and the fluctuation of fractal dimensions 
	of BHH eigenvectors in the interaction basis at different energies from the centre of the spectrum's bulk. However, the agreement between EGOE and BHH in the tunneling basis is more restricted, despite the close matrix structure of both Hamiltonians. 
	
	Scaling for increasing Hilbert space dimension $\NN$
	confirms the convergence to ergodicity of the BHH eigenvector characteristics in the chaotic region, and also exposes the strong basis dependence of such process for certain parameters.
	The scaling analyses suggest that the coefficient of the dominant asymptotic $\ln^{-1}\NN$ correction that governs the convergence of the mean fractal dimensions $\aDq{1,2}$ to $1$, for BHH in the interaction basis and for EGOE, coincides with the one for GOE. 
	Notwithstanding this agreement,
	the three models become more distinguishable from each other in terms of the fractal dimension distributions as Hilbert space dimension grows, in a similar way as we found previously for BHH and GOE \cite{Pausch2020}. 
	This behaviour shows that the distinctive statistics of BHH in the chaotic regime cannot be traced back solely to the two-body nature of the Hamiltonian and the associated Fock-space sparsity, but must be connected to specific correlations among the Hamiltonian's matrix elements. 
	
	These results provide further evidence of a way to discriminate among different many-body Hamiltonians in the chaotic regime, where coarse-grained dynamics
	are governed by common universal features. 
		
	\begin{acknowledgments}
	The authors thank A.~Ortega for fruitful discussions, and acknowledge support by the state of Baden-W\"urttemberg through bwHPC and the German Research Foundation (DFG) through Grants No.~INST 40/467-1 FUGG (JUSTUS cluster), No.~INST 40/575-1 FUGG (JUSTUS 2 cluster), and No.~402552777.
	E.G.C.~acknowledges support from the Georg H.~Endress foundation.
	A.R.~acknowledges support by Universidad de Salamanca through a Grant USAL-C1 (No.~18K145) and by Spanish MCIN/AEI/10.13039/501100011033 through Grant No.~PID2020-114830GB-I00.
	\end{acknowledgments}
	
	\appendix
	
	\section{Further properties of the embedded ensemble. Dependence on $\boldsymbol{\lambda}$ and evaluation of $\boldsymbol{\eps_\text{EGOE}}$}
	\label{sec:EGOEvsLambda}
	\begin{figure}
		\includegraphics[width=.94\linewidth]{\figdir/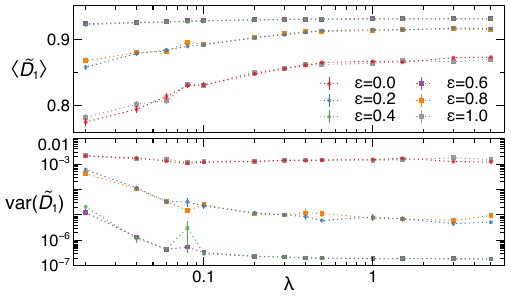}
		\caption{Mean and variance of $\Dq{1}$ versus tuning parameter $\lambda$ for 100 eigenstates of EGOE [Eq.~\eqref{eq:EGOEdef}] around the specified target energies $\eps$, averaged over 100 realizations. Error bars are defined solely from ensemble average.}
		\label{fig:appendix_EGOEvsLambda}
	\end{figure}
	\begin{figure}
		\includegraphics[width=.94\linewidth]{\figdir/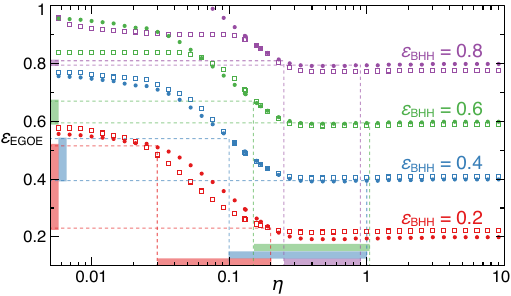}
		\caption{Scaled energy $\eps_\text{EGOE}$ as a function of $\eta$ for four values of $\eps_\text{BHH}$, according to Eq.~\eqref{eq:EGOEmaxdos} (filled circles), 
		and alternatively derived from the condition to bound the same percentage of the total spectrum from above (squares). 
		Dashed lines and thick shades on the axes highlight the $\eta$ intervals associated with the BHH chaotic regions in Fig.~\ref{fig:Dq_vs_eta_allEps} and the corresponding $\eps_\text{EGOE}$ ranges.}
		\label{fig:appendix_epsEGOE}
	\end{figure}
	The emergence of the chaotic phase in Hamiltonian \eqref{eq:EGOEdef} as a function of $\lambda$ is depicted in Fig.~\ref{fig:appendix_EGOEvsLambda} in terms of $\aDq{1}$ and $\vDq{1}$, 
	obtained from 100 eigenstates for different $\eps$ and averaged over 100 ensemble realizations. 
	While for $\lambda\lesssim 0.1$, $\aDq{1}$ clearly increases and $\vDq{1}$ decreases with $\lambda$, both quantities essentially converge to a constant value for $\lambda\gtrsim0.5$ and all $\eps$ considered. We have checked that this holds true also for $\Dq{2,\infty}$. Hence our choice of $\lambda=1$ in the main text.
	Furthermore, the behaviour observed for $\eps$ and $1-\eps$ is very similar, in agreement with the symmetry of the ensemble averaged properties of EGOE around $\eps=0.5$ according to its definition, Eqs~\eqref{eq:EGOEdef}--\eqref{eq:EGOEdef2} [see also Fig.~\ref{fig:EE_DOS_r_Dqmeanvarskew}].
	
	\label{sec:epsEGOE} 
	The chaotic properties of BHH and EGOE cannot
	be compared
	around the same $\eps$ value, due to the shift of the bulk of the spectrum for BHH with $\eta$. 
	In order to enable the comparison, we determine a correspondence between $\eps_\text{BHH}$ and $\eps_\text{EGOE}$ in two different ways: First, one may demand that
	$\eps_\text{BHH}$ and $\eps_\text{EGOE}$ be located at the same distance from the corresponding bulk centre, which we identify with the position of the DOS maximum, $\eps^*_\text{BHH} (\eta)$ and 
	$\eps^*_\text{EGOE}=0.5$, respectively. Then, for chosen $\eps_\text{BHH}$ and $\eta$, we have 
	\begin{align}
		\eps_\text{EGOE}(\eps_\text{BHH},\eta) = \eps_\text{BHH}-\eps^*_\text{BHH}(\eta) + 0.5.
		\label{eq:EGOEmaxdos}
	\end{align}
	Alternatively, the energy equivalence could be established by requiring that $\eps_\text{EGOE}$ and $\eps_\text{BHH}$ have the same position in terms of percentage of the spectrum, i.e., that they bound from above the same fraction of the associated spectra.  
	Figure \ref{fig:appendix_epsEGOE} shows the energy $\eps_\text{EGOE}$ as a function of $\eta$ for four different $\eps_\text{BHH}$ values according to the two methods described, highlighting the ranges of $\eps_\text{EGOE}$ corresponding to the $\eta$ intervals associated with the chaotic region of BHH in Fig.~\ref{fig:Dq_vs_eta_allEps}.
	Even though the two definitions of $\eps_\text{EGOE}$ differ for some $\eta$, they both yield a very similar $\eps_\text{EGOE}$ interval for the chaotic region of BHH at the considered $\eps_\text{BHH}$. 
	Note that due to the discussed symmetry of EGOE, the $\eps_\text{EGOE}$ intervals for $\eps_\text{BHH}=0.2$ and $0.4$ can be truncated at 0.5.

%


\begin{thebibliography}{115}
\bibitem [{MyN()}]{MyNote2}%
	\BibitemOpen
	\href@noop {} {}\bibinfo {note} {For PBCs and odd $L$, where the
		single-particle spectrum is not symmetric around $E=0$ in contrast to PBCs
		with even $L$ and to HWBCs, the spectrum is almost nondegenerate around
		$\eps$=0.5 and consequently the fractal dimensions in the tunneling basis
		denote almost full localization \cite{pausch}.}\BibitemShut {Stop}%
\end{thebibliography}
\end{document}